\documentclass[aps,epsfig,showpacs]{revtex4}

\usepackage{graphicx}

\begin{document}
 
\title{Self-consistent approach to deformation of intruder states 
 in neutron-deficient Pb and Po} 
 
\author{N.~A.~Smirnova$^{1,2}$, 
          P.-H.~Heenen$^3$, 
           G.~Neyens$^1$ } 
 
\affiliation{$^{1}$ Instituut voor Kern- en Stralingsfysica, University of 
Leuven, Celestijnenlaan 200 D, B-3001 Leuven, Belgium} 
\affiliation{$^{2}$ 
Vakgroep Subatomaire en Stralingsfysica, Universiteit Gent, 
Proeftuinstraat 86, B-9000 Gent, Belgium} 
\affiliation{$^{3}$ Service de Physique Nucl\'eaire Th\'eorique, Universit\'e 
Libre de Bruxelles, 
C.P. 229, B-1050 Bruxelles, Belgium} 
 
\date{\today}

\begin{abstract} 
 We present systematic calculations of the properties of $11^-$ isomers 
 in neutron-deficient $^{184-198}$Pb and $^{188-200}$Po. These states are 
based on the $\pi (h_{9/2}i_{13/2})_{K^{\pi }=11^-}$ configuration. They 
are calculated  in the framework of the Hartree-Fock-Bogoliubov method 
 with a Skyrme interaction and density-dependent pairing force. 
The energies and deformations of the $11^-$ states are compared to  
those of the intruder $0^+$ states in both Pb and Po isotopes.   
 In the most neutron-deficient Po isotopes, the calculations predict, 
 below a weakly oblate $11^-$ state, another oblate $11^-$ state 
 which is even more deformed than the intruder oblate $11^-$ state in their Pb isotones. 
The energies and quadrupole moments of the $11^-$ isomers, 
corresponding to a weakly oblate nuclear shape, are in fair 
agreement with the available experimental data.

\end{abstract} 
 

\maketitle 
 
 

The neutron-deficient nuclei around the $Z=82$ shell closure 
constitute one of the most famous examples of shape 
coexistence~\cite{Heyde83,Wood92}. 
In Pb isotopes, oblate, prolate and superdeformed configurations 
coexisting with a spherical ground state were predicted theoretically 
(see~\cite{AFN90} and references therein) and identified 
experimentally~\cite{Wood92,JK91} more than ten years ago. 
By now, low-lying  $0^+$ states, which are a characteristic feature of shape 
coexistence, are known from $^{202}$Pb down to 
$^{184}$Pb~\cite{VDCo,BijnAnd,Andrei00,AndreiEPJ99}. 
The interpretation of these low-lying $0^+$ states relies on proton excitations across 
the $Z=82$ spherical shell gap. In  a shell model 
picture~\cite{HeJo87,HeVI94}, the energy of these 
multi-particle-multi-hole (mp-mh) configurations is lowered by the 
quadrupole-quadrupole and pairing interactions. In mean-field 
models~\cite{BeNa89,Naz93}, these structures are described by a 
state with an oblate (prolate) quadrupole deformation, whose low 
energy is due to the occupation of deformed intruder orbits, such 
as the down-sloping $\pi h_{9/2}$-orbital for an oblate state.

High-spin isomers constructed on these $0^+$ states are of 
particular interest since the measurement of their static 
electromagnetic moments provides direct information on both their 
single-particle structure and their deformation. 
Several isomeric states are known in Pb isotopes. 
In particular, $11^-$ levels have been 
observed from $^{188}$Pb to $^{198}$Pb (see, e.g. Ref.~\cite{JuHe01}). 
They are not known in heavier Pb, most probably because they are not yrast. 
These states are interpreted as a two-quasiparticle ({\it 2qp}) 
excitation based on the $\pi (h_{9/2}i_{13/2})_{K^{\pi }=11^-}$ 
single-particle configuration as confirmed in $^{196}$Pb by the 
measurement of its $g$-factor~\cite{PeHe87}. 
Experimental information on deformation is extracted from 
the collective bands based on the excited 
$0^+$~\cite{VDCo,PeHe87,Andrei00,DrBy02} and  $11^-$ states~\cite{ClWa93,DrBy98}. 
Recent data on static quadrupole moments for $11^-$ states in $^{194,196}$Pb 
~\cite{VyCh02,VyOr02}  support the mean-field interpretation 
of these states as weakly deformed oblate configurations.

The experimental excitation energies of the $11^-$ states and of the $0^+_2$ states 
in Pb isotopes are shown in the left part of Fig.~1. 
Both exhibit a similar behavior. 
The steady decrease of energy with decreasing neutron number 
indicates that, in a shell model picture, 
the residual quadrupole and pairing interactions, or, 
in mean-field models, the deformation effects, are maximal at the 
middle of the neutron sub-shell.

Similar coexisting structures based on intruder configurations are 
known experimentally above and below the $Z=82$ shell closure, in 
the neutron-deficient Hg, Pt and Po isotopes~\cite{Wood92,JuHe01}. 
The energy systematics of the low-energy excited $0^+$ states in 
neutron-deficient $^{200-196}$Po ~\cite{BijnPRL,JuHe01} strongly 
resembles that in Pb. The observation of bands based on these 
states seems to indicate a strong mixing between different shapes. 
The $11^-$ isomers have been identified from the semi-magic 
$^{210}$Po down to $^{194}$Po~\cite{JuHe01}. In contrast to the Pb 
case, their excitation energy is nearly constant as a function of 
the neutron number.

Previous theoretical studies of these isotopes have been mainly 
devoted to the description of the $0^+$ intruder states, in 
particular in Pb. A large variety of models has been used: a 
schematic spherical shell model~\cite{VDCo,HeJo87,DCDe00} and its 
symmetry-based boson truncations~\cite{FoHe02}, the Strutinsky 
approach with a Woods-Saxon potential~\cite{MaPa77,BeNa89,Naz93}, 
and several variants of self-consistent mean-field 
methods~\cite{TaFl93,ChEg01,HeVa01,DuBe02}. A comprehensive study 
of the low-energy spectroscopy and ground-state properties of 
light Po isotopes has been presented within several schematic 
approaches in Ref.~\cite{OrHe99} (see review ~\cite{JuHe01} for 
more references). On the other hand, the only systematic study of 
isomers in these neutron-deficient nuclei is based on cranked 
Nilsson-Strutinsky type calculations with a schematic Woods-Saxon 
potential and is devoted to $^{186-200}$Pb~\cite{BeNa89,ClWa93}. 
The quadrupole moments of $11^-$ intruder states in Pb and Po down 
to  $^{194}$Pb and $^{198}$Po have also been analyzed recently 
within the particle-core coupling model~\cite{VyOr02}. The 
deformation properties of the $11^-$ state in $^{196}$Pb have been 
discussed in the framework of the tilted-axis cranking 
model~\cite{VyCh02}. The magnitude of the $11^-$ nuclear 
deformation is indeed a basic ingredient in this model for the 
description of magnetic rotational bands in both even-$A$ and 
odd-$A$ neutron-deficient Pb isotopes~\cite{ClMa00,VyCh02}. 
However, up to now, no fully microscopic investigation of these 
isomeric configurations has been performed.

In this letter, we present the results of the first fully 
self-consistent description of the properties of $11^-$ intruder 
isomers in $^{184-198}$Pb and in $^{188-200}$Po based on the 
Hartree-Fock-Bogoliubov (HFB) method. Our aim is to analyze 
the nature of these states and to study the evolution of their 
deformation as a function of neutron number. The simultaneous 
consideration of Pb and Po isotopes will allow also to determine 
the effect of the $Z=82$ spherical shell closure on the 
deformation trends. Since several Po isotopes have extremely 
shallow energy surfaces as a function of the axial quadrupole 
moment, we have also performed selected configuration-mixing 
calculations to validate the construction of isomers on an oblate 
configuration. 
 
All our calculations have been performed with the same set of nucleon-nucleon 
effective interactions.  In the mean-field channel, we use the Skyrme 
parameterization Sly4~\cite{Sly4} and in  the pairing channel, 
a density-dependent zero-range force mainly active on the nuclear surface 
as introduced in Ref.~\cite{RiBo99}. 
The HFB equations have been solved on a three-dimensional cubic 
mesh~\cite{TeHe96} for reflection symmetric nuclear shapes with the method 
presented in  Refs.~\cite{GaBo94,TeHe96} and include 
an approximate particle-number projection with the Lipkin-Nogami prescription~\cite{LN}. 
The isomeric states are  constructed fully self-consistently 
as $2qp$ excitations as in Ref.~\cite{HeJa98}, 
selecting the quasiparticles whose dominant components 
in the HF basis are the $9/2^-$ and  $13/2^+$ particle states. 
 
In Pb isotopes, we have taken for the energy and the deformation of the first 
excited $0^+$ states those of the oblate minimum 
of the deformation energy curves. Configuration-mixing 
calculations~\cite{TaFl93,ChEg01,HeVa01,DuBe02} have shown that 
this approximation is well justified, even when the potential well 
of the oblate minimum is very shallow. The topology of the energy 
curves for Po isotopes is more complex, with several shallow 
minima on both prolate and oblate sides. The properties of the 
first excited $0^+$ state have therefore been determined by 
performing configuration-mixing calculations with respect to the 
axial quadrupole collective coordinate. To avoid the ambiguity 
resulting from the fact that the states obtained from the 
configuration-mixing calculations do not have the same 
mean-particle number, we have included in the configuration mixing a 
projection on the correct proton and neutron numbers. 
 
This set of methods has already been applied with success to 
several other problems, such as the description of neutron-rich 
nuclei~\cite{TeHe96,TeFl96}, of superdeformed rotational 
bands~\cite{TeFl97,HeJa98} and of heavy nuclei~\cite{DuBo01}. \\

%
{\it Deformation energy curves}\\ 
The deformation energy curves obtained for $^{194-184}$Pb and 
$^{198-188}$Po isotopes are shown in Figs.~2 and 3, respectively. 
These curves have 
been obtained by HF+BCS calculations with a constraint on the 
axial mass quadrupole moment $q\equiv q_{20}=\langle 2 
z^2-x^2-y^2\rangle$.  Our formalism for configuration mixing 
calculations is indeed up to now limited to the mixing 
of HF+BCS wave functions and does not permit to use HFB ones. 
For this reason, we shall base 
our discussion of the deformation properties on the HF+BCS energy 
curves. We have checked that 
very similar curves are obtained within the 
constrained HFB method and 
that the trend in the $0^+_2$ excitation energy as a 
function of the neutron number is the same. In order to make 
a connection with simpler models, we have labelled orbitals by 
approximate Nilsson indices. 
 
For all Pb isotopes (Fig.~2), the ground state corresponds to the spherical 
minimum. With decreasing neutron number, two structures appear, 
first as shoulders, then as secondary  minima on the oblate 
(starting from $A=192$) and prolate sides (starting from $A=188$). 
The oblate minimum occurs around $q=-10$ b which is the 
deformation corresponding to the crossing between the down-sloping 
$9/2[514]$ mean-field orbital  and the up-going orbital $1/2[440]$. 
On the prolate side, the minimum at $q \sim 20 $ b 
corresponds to the deformed shell gap due to the $1/2[530]$, 
$1/2[541]$ and $3/2[532]$ intruder orbitals and the rising of the 
$1/2[400]$, $3/2[402]$ and  $11/2[505]$ orbitals. 
For $^{186}$Pb, the excitation energy of the prolate minimum significantly 
overestimates the experimental data. 
This discrepancy is irrelevant for the description of the 
isomeric states which are constructed on the oblate minimum. 
The energy of this prolate state is very sensitive to 
the details of the mean-field method. 
Several other mean-field calculations, 
based either on the HF+BCS method and other Skyrme 
interactions~\cite{TaFl93,DuBe02}, or on the HFB 
method with the Gogny force~\cite{ChEg01}, predict a prolate 
minimum lower in energy than the oblate one.

As we have explained in the introduction, for the Pb isotopes, 
we have approximated the energy of the oblate 
$0^+$ state by the energy of the oblate minima. 
These energies  are compared with the experimental data 
in the left part of Fig.~1.  The decrease of the 
HF+BCS energies with decreasing neutron number is 
in good agreement with the data, 
although the experimental energies are overestimated. 
This overestimation is even larger  by 
approximately 300 keV in HFB calculations. 
 
The  deformation energy curves for Po isotopes (Fig.~3) exhibit a 
complicated coexistence of several shallow minima. To check whether 
these minima are stable against triaxial deformations, we have 
performed selected calculations including constraints on the 
triaxial quadrupole moment. These calculations show that the 
symmetric minima corresponding to $q \sim \pm 8$ b ($\beta \sim 
\pm 0.1$) obtained for all Po isotopes are separated by a very low 
barrier of about 100 keV in the $\gamma $ plane. They are 
therefore probably unstable against $\gamma $ vibrations. 
The shoulder that appears in  $^{198}$Po for a deformation around 
$-15$ b becomes a deep well for lighter isotopes which is 
the ground state for $^{192,190}$Po. 
In $^{188}$Po, the oblate minimum is nearly degenerate in energy 
with a prolate minimum at $q\sim 19$ b. The Strutinsky  calculations of 
Refs.~\cite{MaPa77,BeNa89,OrHe99} agree qualitatively with our 
results. It is clear from this discussion that a 
study of the low lying levels of Po isotopes must 
take into account the vibrational motion with 
respect to the axial quadrupole mode.  \\

%
{\it Properties of low-lying oblate $0^+$ states: configuration-mixing 
calculations for Po isotopes} \\ 
We have performed configuration-mixing calculations with respect 
to the axial quadrupole moment for the Po isotopes within
the generator coordinate method (GCM), 
as was presented in Refs.~\cite{BoDo90,HeBo93}. 
The basis for the calculations is a set of wave functions $\Phi (q)$ 
obtained by  HF+BCS calculations with  a constraint on the axial 
quadrupole moment. These wave functions $\Phi (q)$ are projected 
on the correct number of protons and neutrons. Collective wave 
functions are then constructed as a linear superposition of the 
projected non-orthogonal basis functions. The weights 
corresponding to the different quadrupole moments are determined 
by diagonalizing a discretized version of the Hill and Wheeler 
equation~\cite{HiWh53,BHR03}. This procedure provides as many 
eigenstates as discretization points of the quadrupole moment. The 
lowest eigenvalue corresponds to the ground state with a wave 
function spread over the axial quadrupole moment. As we shall see, 
depending on the structure of this wave function, it represents either 
a vibration within a well-defined well or a mixing of mean-field 
states with different shapes. In the same way, the other 
eigenvalues of the Hill and Wheeler equation represent excited states.  
Depending upon their shape, they can be classified into three main categories. 
Some states are mainly located in a well that is clearly separated 
from the ground state (as it is the case for a super-deformed 
well). Other states can be $\beta$ vibrations with respect to a 
lower state in the same potential well. Finally, some states can 
represent a mixing of states corresponding to different shapes. 
We have not included triaxial deformations in these configuration-mixing 
calculations. Some Po isotopes being soft against 
the $\gamma$ degree of freedom, these deformations may affect some 
of our results. However, their inclusion would considerably increase the 
computation time, and since our main purpose is to calculate the $11^-$ isomers, 
we have limited the configuration mixing to axial shapes.

The  energies of the first few states obtained by the 
diagonalization of the Hill and Wheeler equation are shown 
for selected Po isotopes in the upper part of Fig.~4, 
together with the deformation energy curves. 
The triangles representing the energy of the configuration-mixed states 
are positioned at their average quadrupole moment. The amplitudes 
of the collective eigenfunctions (which are related to the mixing 
coefficients by an integral transformation~\cite{RS80}) are 
plotted in the lower part of the figure for the three lowest 
states obtained for each isotope. The wave functions of these 
states are spread over a large range of quadrupole moments. In 
$^{200,198,196}$Po, the ground state and the first excited state 
are a mixing of an oblate and a prolate configuration, resulting 
in an average small oblate deformation. This mean quadrupole 
moment is close to the one of the first shallow oblate minimum of 
the deformation energy curve. In the three isotopes, the second 
excited $0^+$ state has an  average quadrupole moment $q \sim -10$ b, 
corresponding to a weakly oblate shape. While it 
has only a slightly dominant weight in the ground-state wave 
function of $^{200,198,196}$Po, this oblate configuration becomes 
the ground state in  $^{194,192,190}$Po, with still a spreading 
over a large range of quadrupole moments. 
In $^{190}$Po, it is nearly degenerate with a 
predominantly prolate configuration. 
These results agree well with the 
interpretation of the experimental data that 
the ground states of  $^{200-196}$Po isotopes are mixed but nearly spherical, 
the ground state of $^{194}$Po is mixed, with a large oblate component and that of 
$^{192}$Po is predominantly oblate~\cite{BijnAnd,HeCo99}. 
Experiments also suggest a strong mixing between oblate and 
prolate shapes in the $^{190}$Po ground state (\cite{JuHe01,Po190}).

 In the right part of Fig.~1, we compare 
the three first eigenstates of the 
configuration-mixing calculation with the available experimental data. 
The states that have the largest oblate mean deformation are indicated 
by filled triangles. 
The experimental excited $0^+$ states are known in 
$^{200,198,196}$Po~\cite{BijnPRL} and 
are interpreted as 4p-2h excitations in the spherical shell model 
(see Refs.~\cite{BijnPRL,HeCo99,JuHe01} and references therein) 
or as states with an oblate deformation in mean-field models. 
The experimental trend  is well 
reproduced by the most oblate states, although it is not the first excited state. 
Note however that the experimental energies are intermediate between those 
of the two first excited states. This may indicate that the interaction that we use 
leads to a too large mixing between the deformed configurations. 
Such a result could be modified by the inclusion of triaxial deformations, but in 
a way which can hardly be anticipated. 
 
From the deformation energy curves for the Po isotopes,  
the $11^-$ states can be constructed on either of the two oblate minima. 
The configuration-mixing analysis clearly favors the slightly 
oblate minimum for the heavier isotopes. For the lighter isotopes 
($^{194 - 188}$Po), the mean deformation 
of the configuration-mixed ground state is intermediate between the two. 
For these nuclei, we shall therefore explore whether an 
$11^-$ isomer can be constructed for both deformation regions. \\


%
 
{\it Properties of the $11^-$ isomers}\\ 
The $11^-$ states in the Pb and Po isotopes have been constructed 
as a $2qp$ excitation on an HFB vacuum. 
As in  Ref.~\cite{HeJa98}, the $2qp$ state is calculated fully 
self-consistently, which means that effects such as core 
polarisation due to the time-odd terms of the interaction and the 
conservation of the number of particles or changes of deformation 
are taken into account. Since the single-particle basis, which is 
formed by the eigenstates of the Hartree-Fock Hamiltonian, does 
not diagonalize the $qp$ basis, the $qp$ excitations cannot be 
defined exactly by single-particle states. We have always chosen 
the $2qp$ wave functions which have the largest overlaps with 
the lowest $9/2^-$ and $13/2^+$ single-particle states. In 
practice, these overlaps are always larger than 0.95. 
As in previous calculations of $1qp$ and $2qp$ excited states~\cite{HeJa98,DuBe02}, 
the isomeric state has been obtained axial in all cases, although triaxial 
deformations are allowed by the symmetries of our computer code.

The experimental and theoretical excitation energies of the $11^-$ 
states are compared  in Fig.~1 for the Pb and Po isotopes. In the 
Pb isotopes, the energy of the $11^-$ states decreases with 
decreasing neutron number, both in experiment and theory. The 
experimental trends are quite well reproduced by our calculation, 
although the excitation energy is overestimated by 600 to 800 keV. 
This overestimation is most probably related to a better 
description in the HFB formalism of the vacuum (corresponding to 
the ground state) than of $qp$ excitations. Our calculation shows 
that the excitation energy of the $11^-$ state is the lowest in 
$^{190,188}$Pb and increases in $^{186,184}$Pb for which there are 
no experimental data yet.

For the Po isotopes, the $11^-$ states are known down to $^{194}$Po. 
The isomer has been identified in $^{192}$Po~\cite{HeCo99}, 
but its energy could not be obtained from the data. 
In $^{200 - 196}$Po, a minimum corresponding  to an $11^-$ state could be 
constructed only for small oblate deformations. The theoretical 
excitation energies are in good  agreement with the data. In the case 
of $^{194 - 188}$Po,  $2qp$ states have been obtained for both oblate deformations. 
The isomers corresponding to smaller (larger) absolute values of the 
quadrupole moment are shown by filled (empty) diamonds in Fig.~1 (right part). 
The excitation energies of both solutions are close to each other. 
These states have to be further mixed by a configuration-mixing calculation. 
One should not expect a significant change in the excitation energy of 
the lowest mixed eigenstate, 
the ground state being lowered in a similar way by configuration mixing.

The energies of the $11^-$ states are nearly constant as a 
function of the neutron number and the isomers occur at a lower 
energy than in Pb. The $11^-$ energies do not follow the steep 
decreasing trend as the intruder $0^+$ energies. This suggests 
that the structure of the $11^-$ isomers is rather similar for all 
$Z=84$ isotopes, at least down to $^{196}$Po.  
In lighter Po isotopes, the $11^-$ states built on the most oblate 
minimum become lower in energy than the weakly oblate ones, 
and configuration-mixing calculations 
would be necessary to get more accurate predictions on the  $11^-$
energy and deformation. 

In the terminology of phenomenological models, the constant 
energy of the $11^-$ states in heavier Po isotopes, indicates
that the self-consistency effects due to the $2qp$ excitation do 
not affect strongly the core. 
This is not the case for the Pb isotopes, in which the $Z=82$ 
magic number is broken by the $2qp$ excitation, inducing an 
additional core deformation, as also demonstrated by the trend in the 
intrinsic quadrupole moments.

The calculated intrinsic charge quadrupole moments of the 
$11^-$ states in Pb and Po are plotted in Fig.~5. 
For comparison, there are also given two available data points for 
$^{194,196}$Pb  extracted from the measured spectroscopic quadrupole 
moments in the assumption of the axial symmetry and $K=11$~\cite{VyCh02}. 
The theoretical results are in good agreement with the experimental values. 
The deformation parameters ($\beta _2$, $\beta _4$) defined as in 
Ref.~\cite{CiDo96} vary for Pb isotopes 
from $(-0.15,-0.014)$ for $^{198}$Pb to $(-0.19, -0.018)$ for $^{184}$Pb. 
The quadrupole moments of 
the intruder isomers in the Pb-isotopes are significantly larger 
than those of the $11^-$ states corresponding to the weakly oblate shape
in the heaviest Po isotopes. This reflects the influence of 
core polarisation due to the breaking of the $Z=82$ core in the Pb 
isotopes.
 
 
In lighter Po isotopes, two candidates for an $11^-$ isomer 
have to be mixed by a configuration-mixing calculation. 
It is difficult to predict the quadrupole moment of the mixed state
which could have an intermediate value between those of the two mean-field states.  
 From these results, one can infer a smooth change from a slightly 
deformed oblate shape of $11^-$ states in the heaviest Po to a more deformed 
oblate shape in the lightest. Experimental observation of the 
rotational bands above these isomers or a direct measurement of 
the spectroscopic quadrupole moments would be helpful to clarify 
the situation. For the unambiguous $11^-$ states in the heavier Po 
isotopes, the deformation parameters ($\beta _2$, $\beta _4$) vary 
from $(-0.114, -0.007)$ for $^{200}$Po to $(-0.117, -0.007)$ for $^{196}$Po.
 

In summary, we have, for the first time, applied the HFB method 
with a Skyrme interaction and a realistic pairing force to the 
study of the $11^-$ isomers in a series of neutron-deficient Pb 
and Po nuclei. The isomers have been constructed self-consistently 
as $2qp$ excitations $\pi (h_{9/2^-} i_{13/2^+})_{11^-}$. Their 
main properties are in good agreement with the available 
experimental data. We obtain that the energy of $11^-$ states in 
Pb isotopes reaches minimum around $N=106-108$ and increases in 
lighter isotopes for which there are no data yet. For Po isotopes, 
the $11^-$ energy is predicted to remain roughly constant as a function
of neutron number. The calculations support the assumption of oblate 
deformation of these nuclei in their $11^-$ states, with a steady 
increase towards neutron-deficient isotopes. The deformation 
is more pronounced in Pb than in Po, but in lighter 
Po isotopes the deformation of the $11^-$ states might become of 
the same order or even larger than that in the Pb isotopes. We have 
also studied the properties of the low-lying $0^+$ states in Po 
isotopes within the configuration-mixing approach on the basis of 
HF+BCS wave functions. These calculations show that, in all cases, 
the lowest $0^+$ states do not have a well defined deformation, 
and oblate, spherical and prolate shapes are strongly mixed, both 
in the ground and the excited $0^+$ states.

 
\acknowledgements{We acknowledge discussions with A.~Andreyev, 
 D.~Balabanski, M.~Huyse, K.~Van de Vel and P.~Van~Duppen on experimental 
data and  thank M.~Bender for his help in the GCM calculations. 
N.A.S. thanks Kris Heyde for a careful reading of the manuscript and 
valuable comments. 
This work is supported by the DWTC grant IUAP $\#P5/07$.}

\begin{figure} 
 \includegraphics[height=.25\textheight]{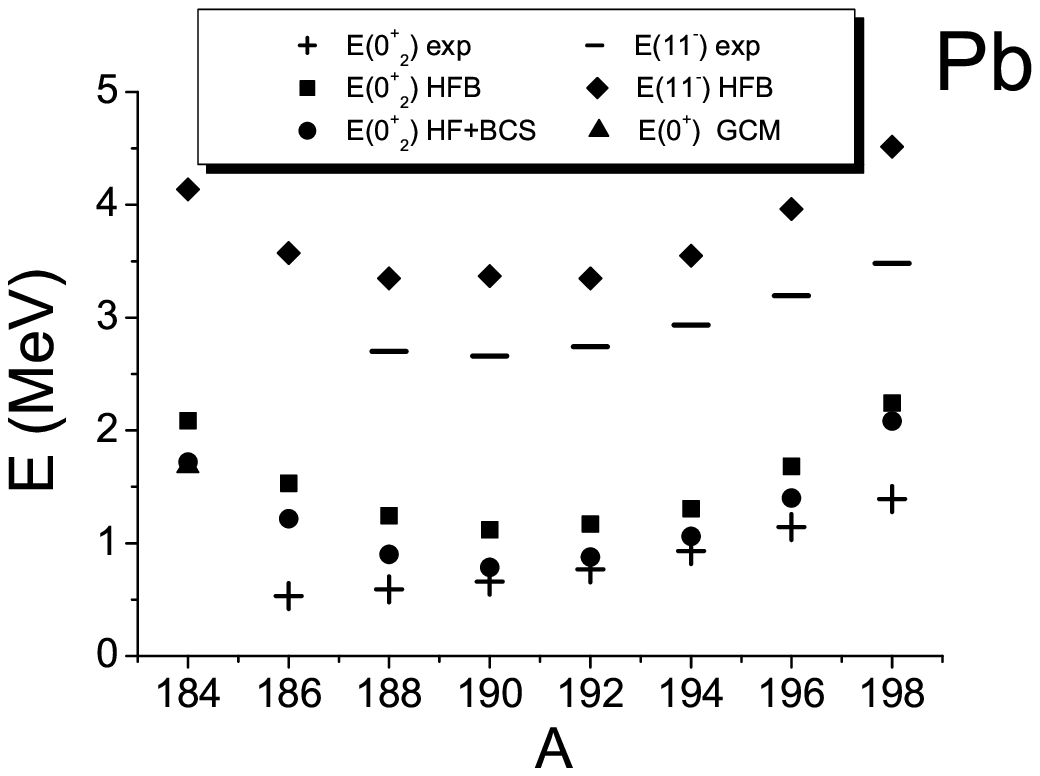} 
 \hspace{5mm} 
 \includegraphics[height=.25\textheight]{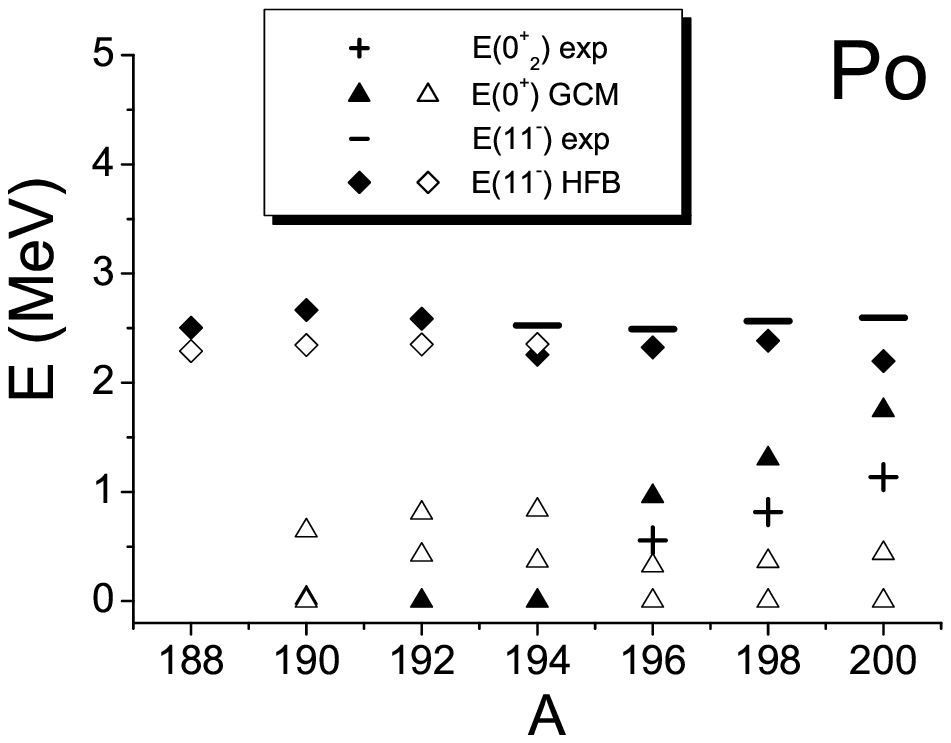} 
 \caption{Theoretical and experimental energies 
of the $11^-$ and $0^+_2$ states in 
 $^{184-198}$Pb (left) and in $^{188-200}$Po (right) 
(see Ref.~\protect\cite{JuHe01} for the references). 
The experimental $0^+_2$ in $^{184}$Pb is not indicated, 
since its nature is not well established~\protect\cite{AndreiEPJ99}. 
For $^{184}$Pb, the oblate $0^+$ state from configuration-mixing calculations
is shown for comparison.
For the theoretical $0^+$ states in Po isotopes, 
the first three eigenvalues obtained in the configuration-mixing 
calculations are represented by triangles. The filled triangles correspond 
to the states with the largest oblate mean deformation. 
The empty diamonds denote the $11^-$ states based on the 
second oblate minima for the cases when convergence was achieved.} 
\end{figure}

\begin{figure} 
 \includegraphics[height=.18\textheight]{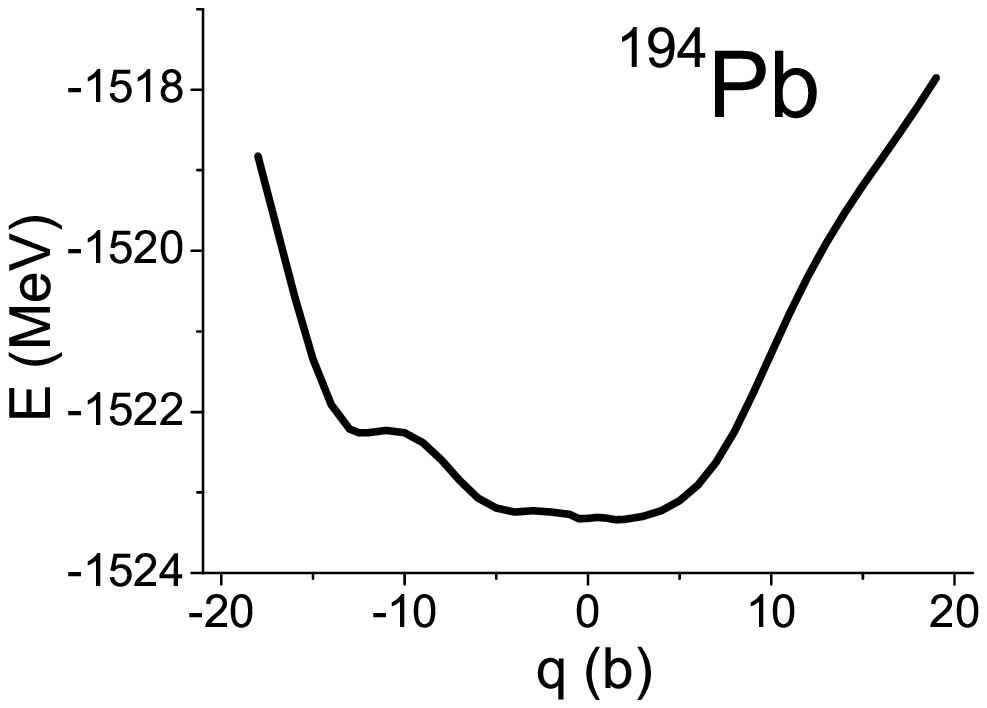} 
 \includegraphics[height=.18\textheight]{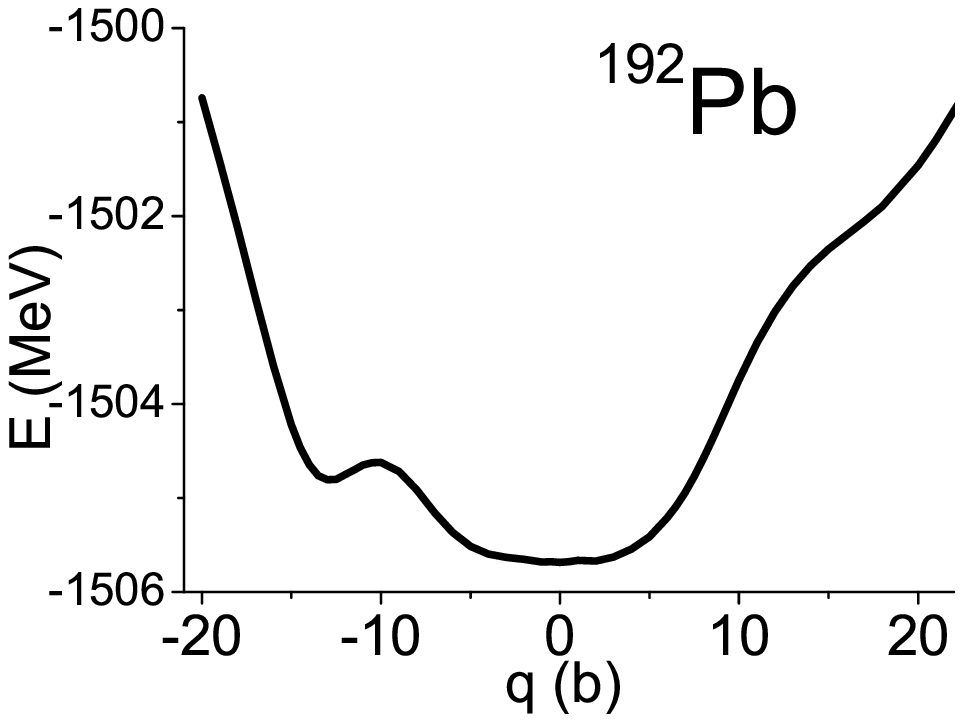} 
 \includegraphics[height=.18\textheight]{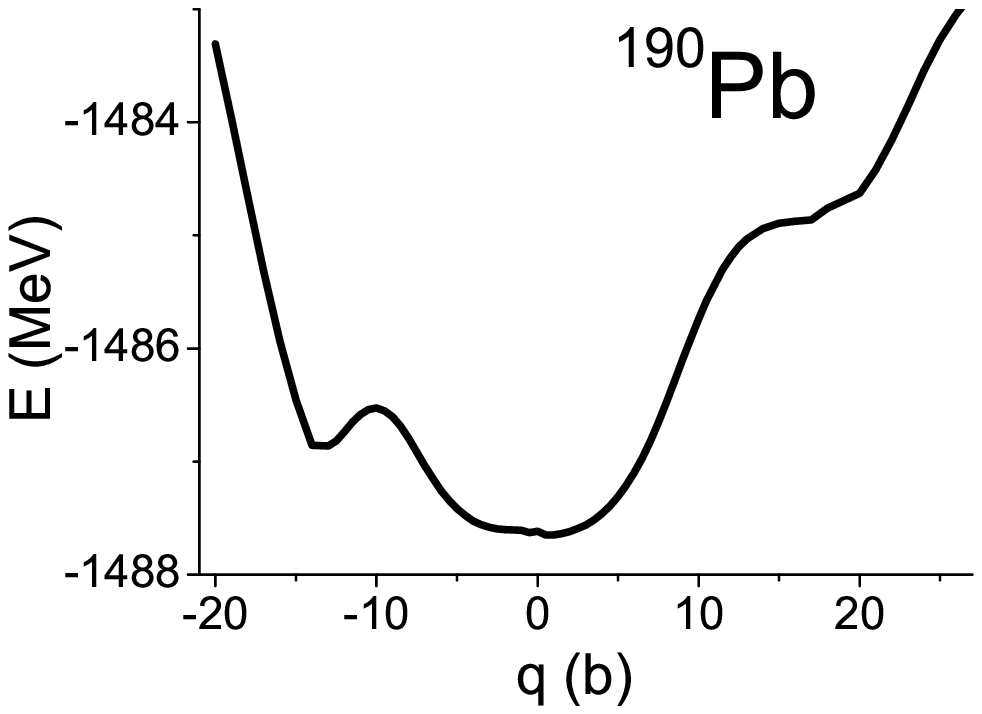}\\ 
 \includegraphics[height=.18\textheight]{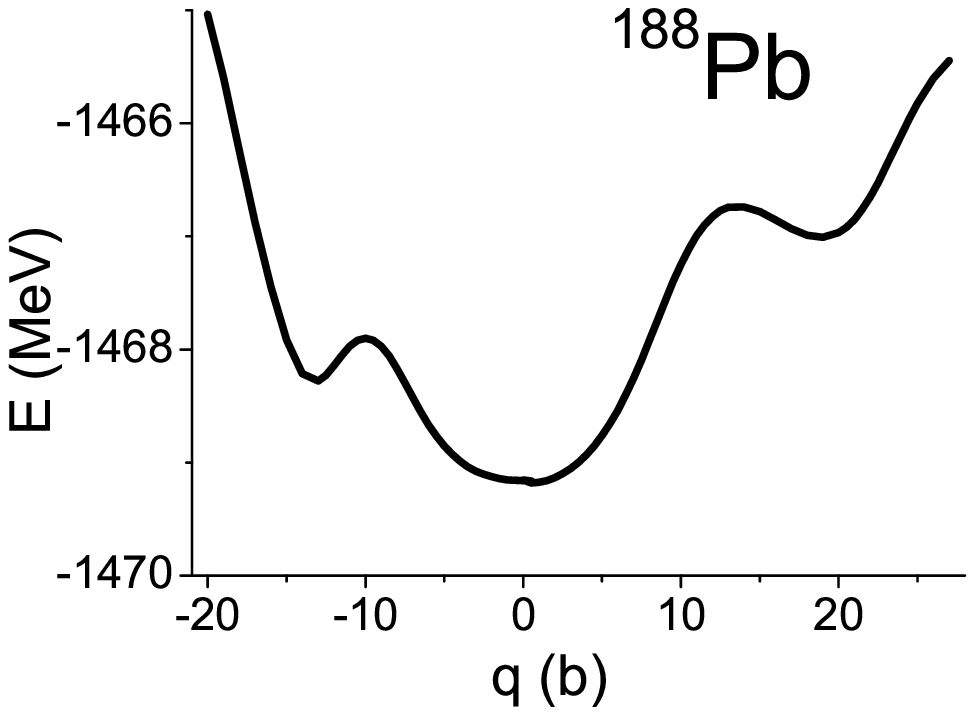} 
 \includegraphics[height=.18\textheight]{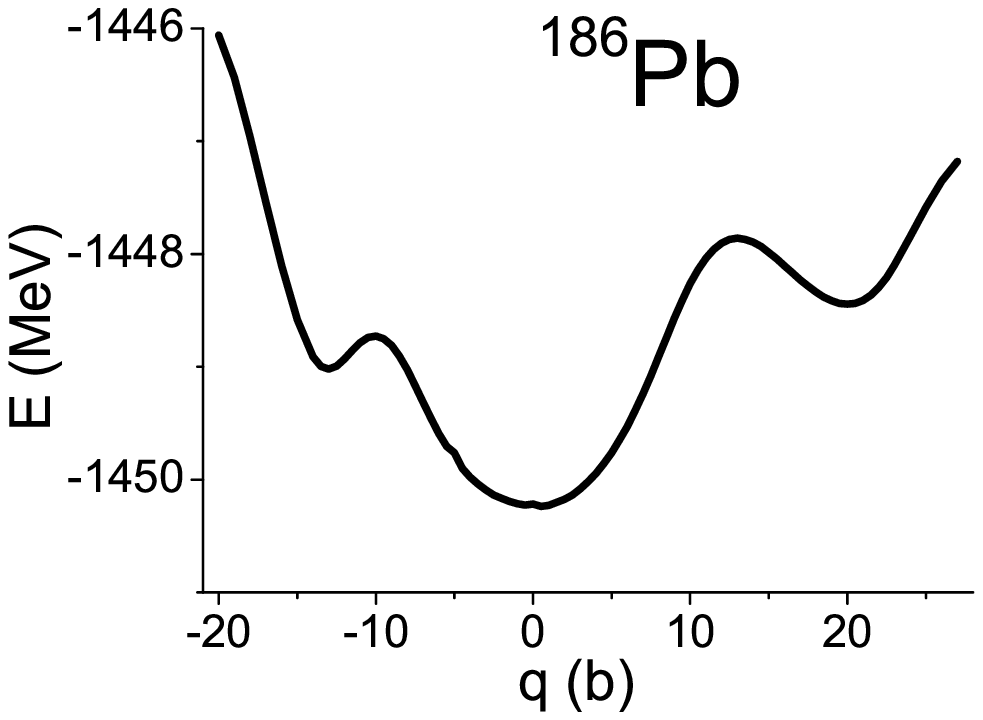} 
 \includegraphics[height=.18\textheight]{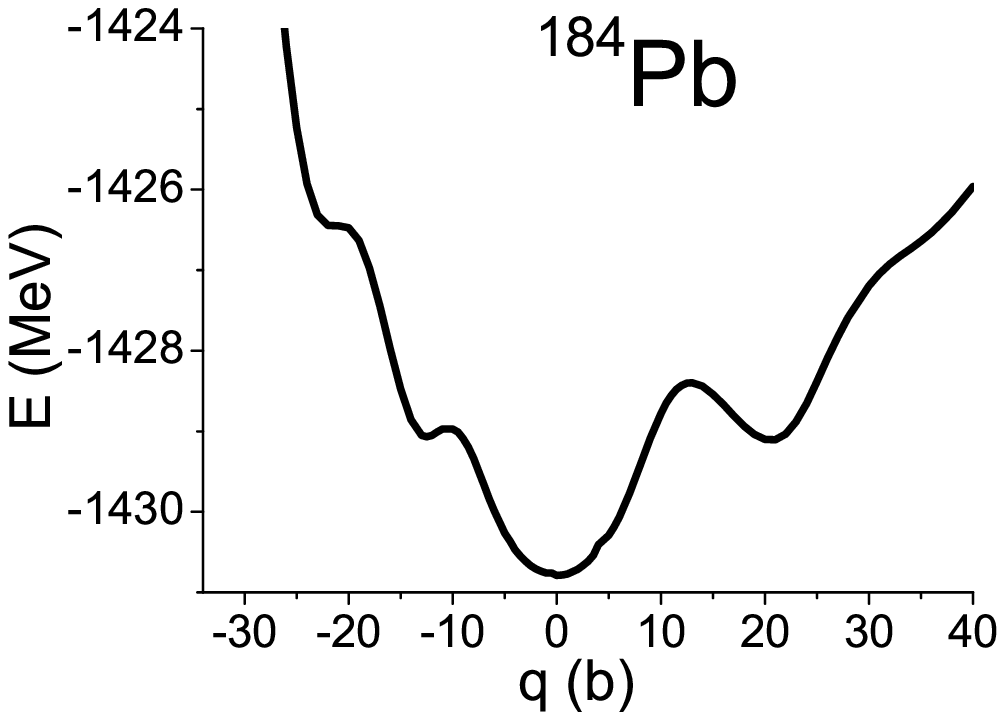} 
 \caption{ Deformation energy curves for $^{194-184}$Pb 
obtained  from the constrained HF+BCS calculations.} 
\end{figure}

\begin{figure} 
 \includegraphics[height=.18\textheight]{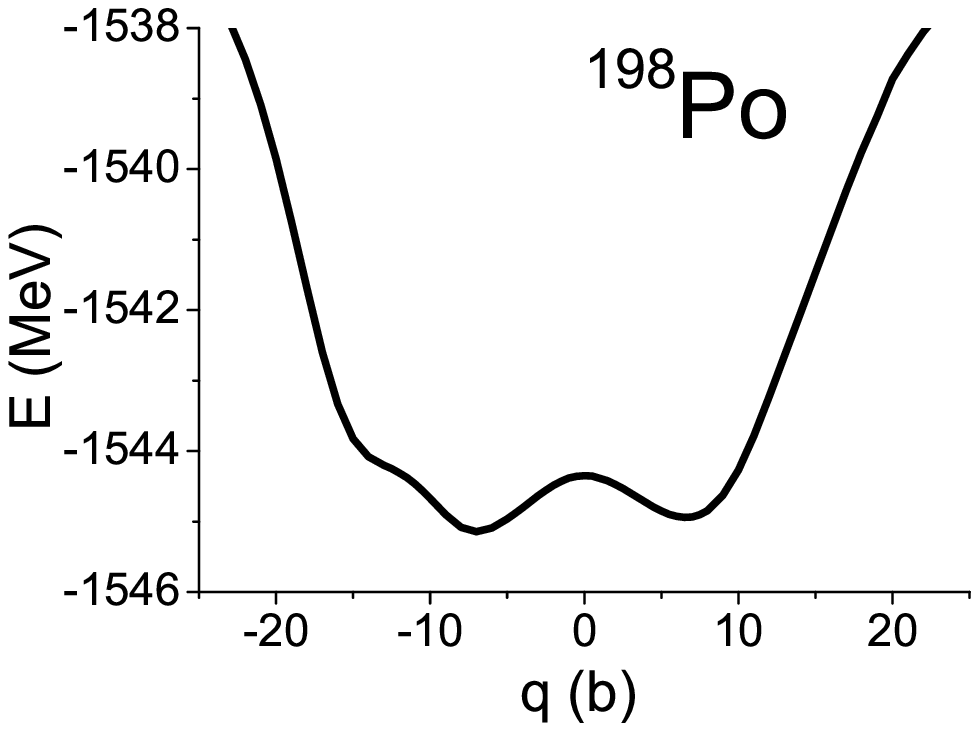} 
 \includegraphics[height=.18\textheight]{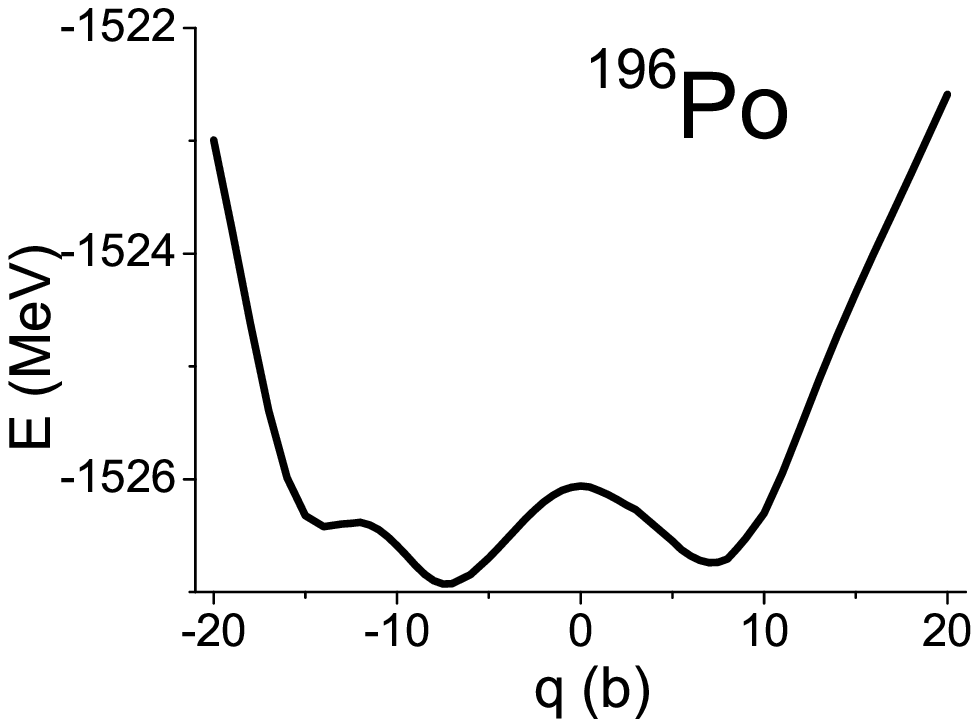} 
 \includegraphics[height=.18\textheight]{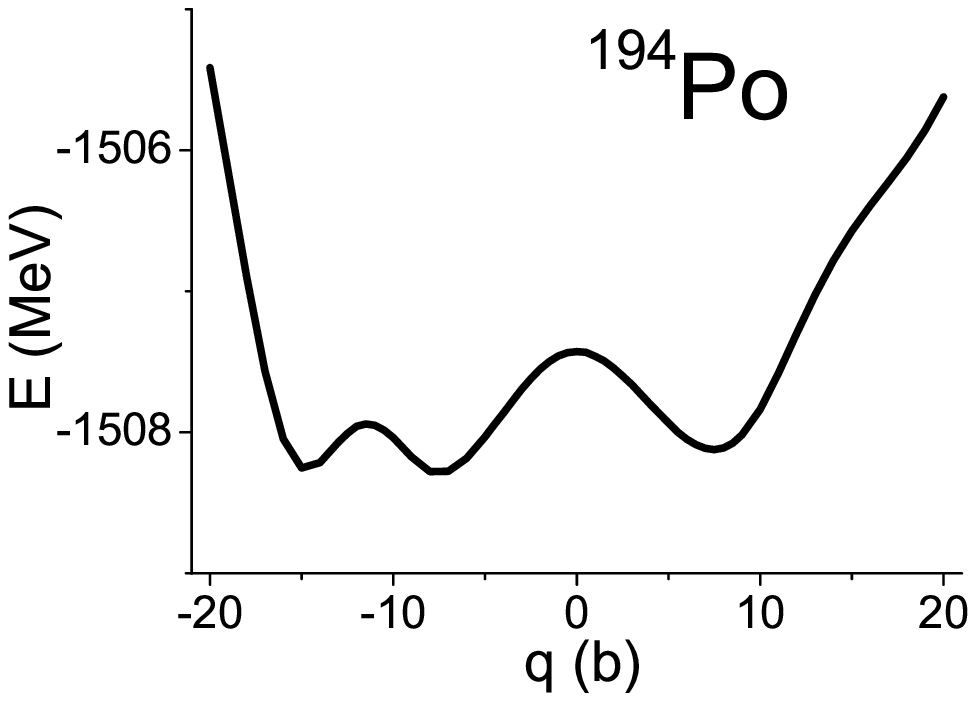}\\ 
 \includegraphics[height=.18\textheight]{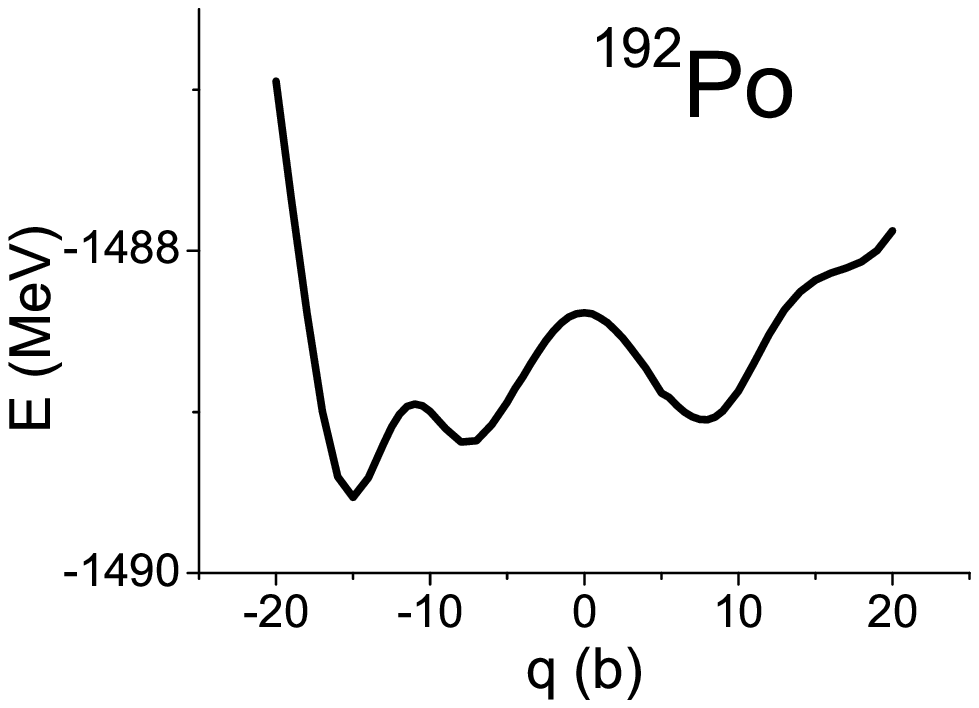} 
 \includegraphics[height=.18\textheight]{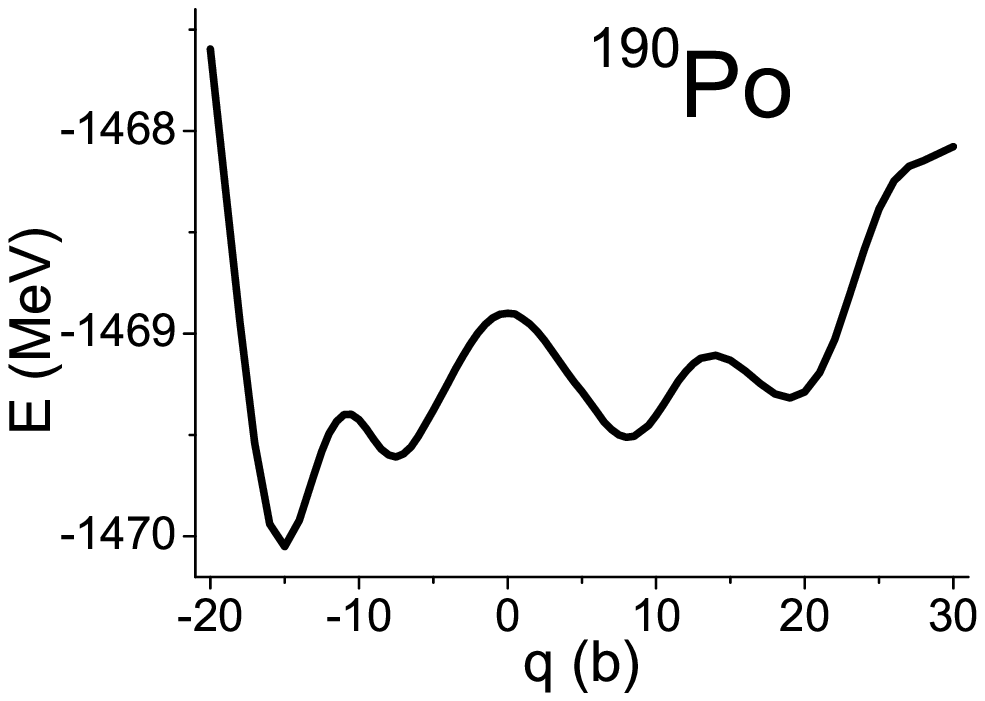} 
 \includegraphics[height=.18\textheight]{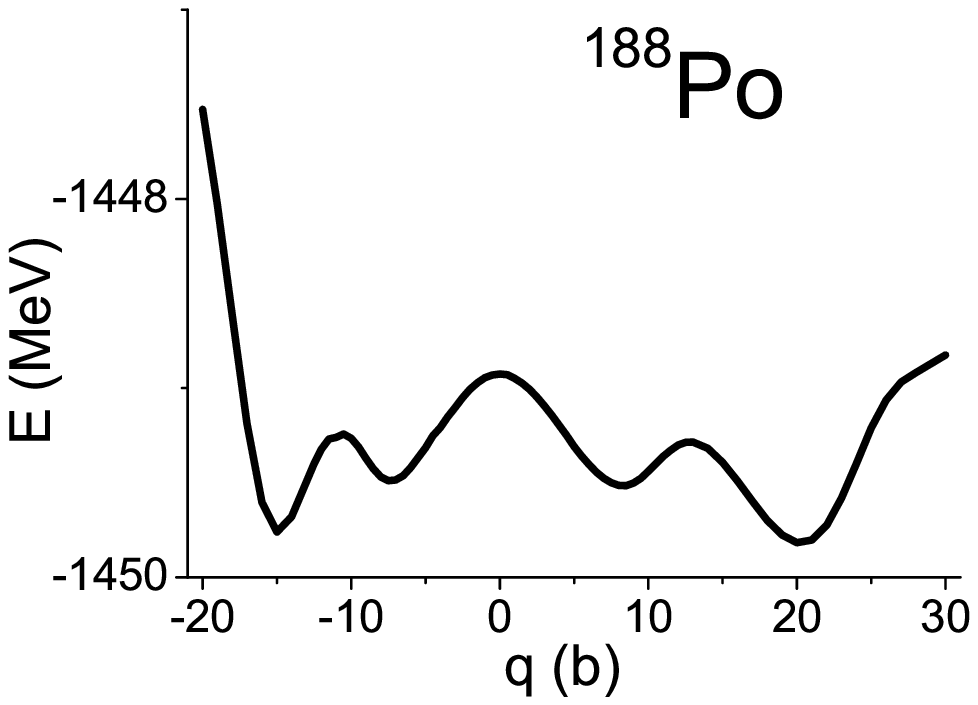} 
 \caption{ Deformation energy curves for 
 $^{198-188}$Po obtained from the constrained HF+BCS calculations.} 
\end{figure}

\begin{figure} 
 
\begin{tabular}{ccccc} 
 \includegraphics[height=.12\textheight]{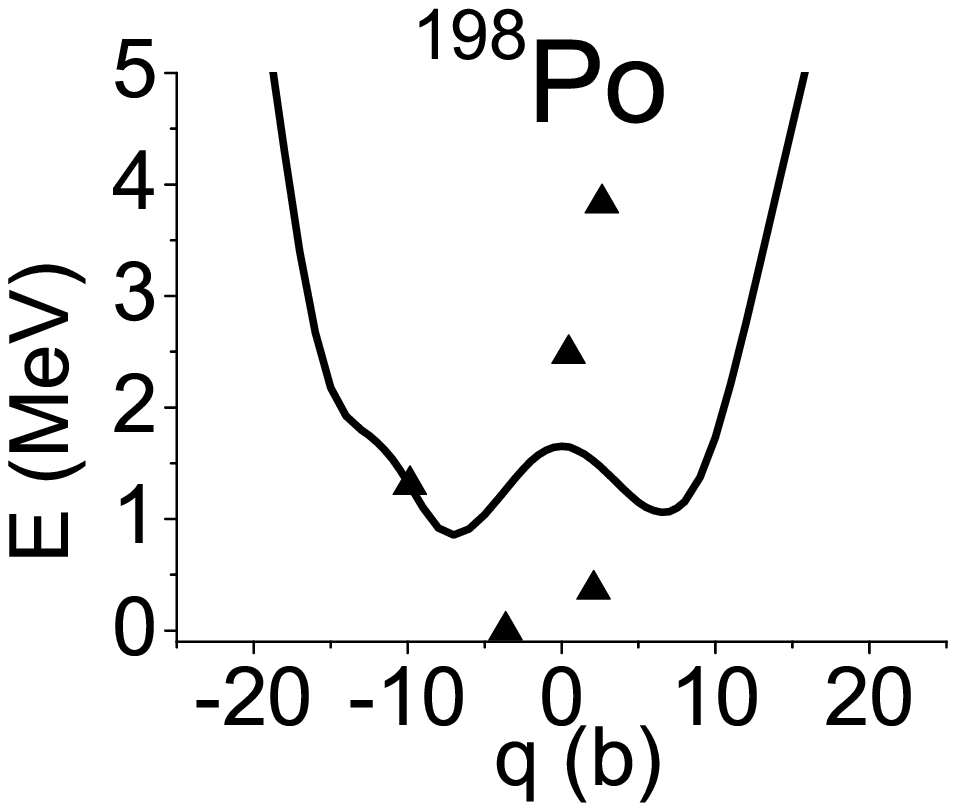} & 
 \includegraphics[height=.12\textheight]{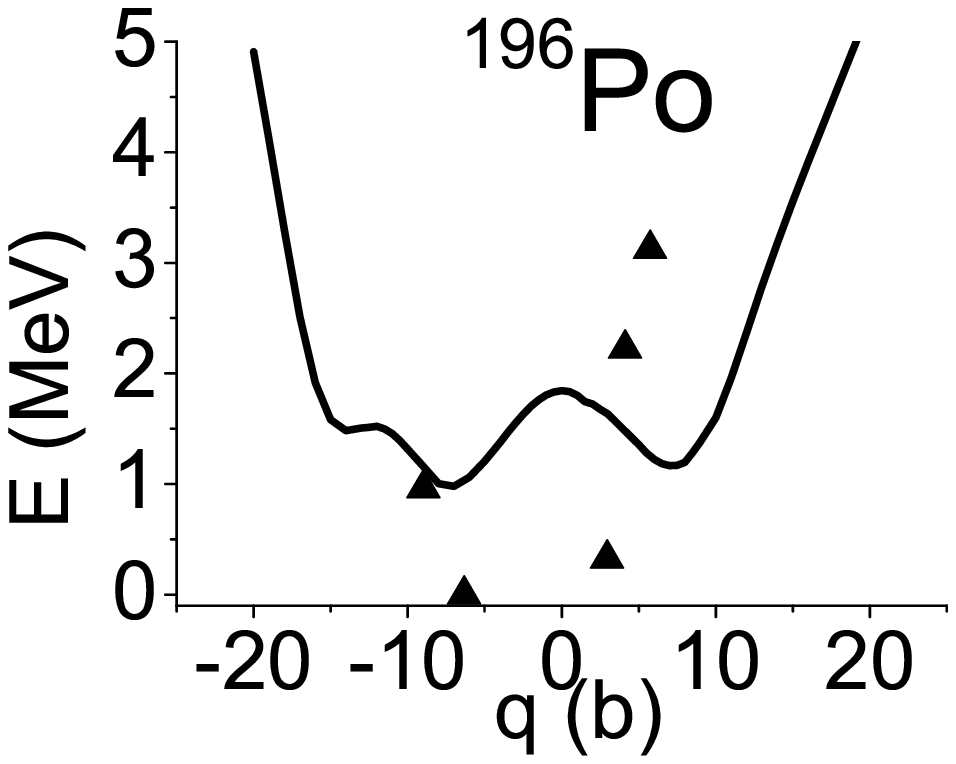} & 
 \includegraphics[height=.12\textheight]{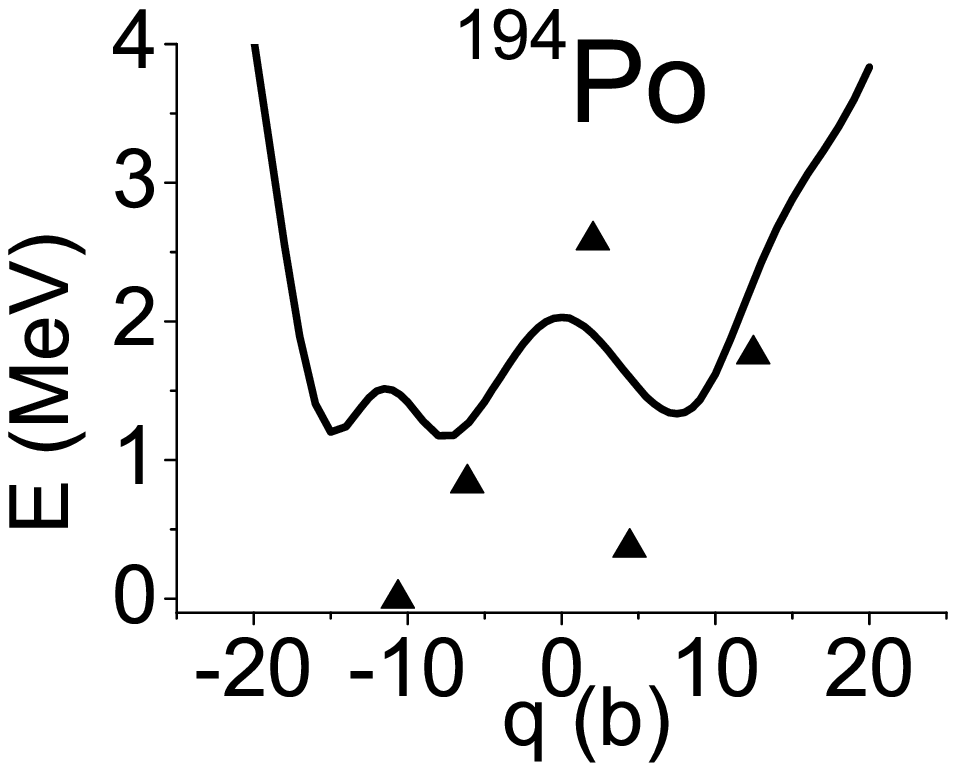} & 
 \includegraphics[height=.12\textheight]{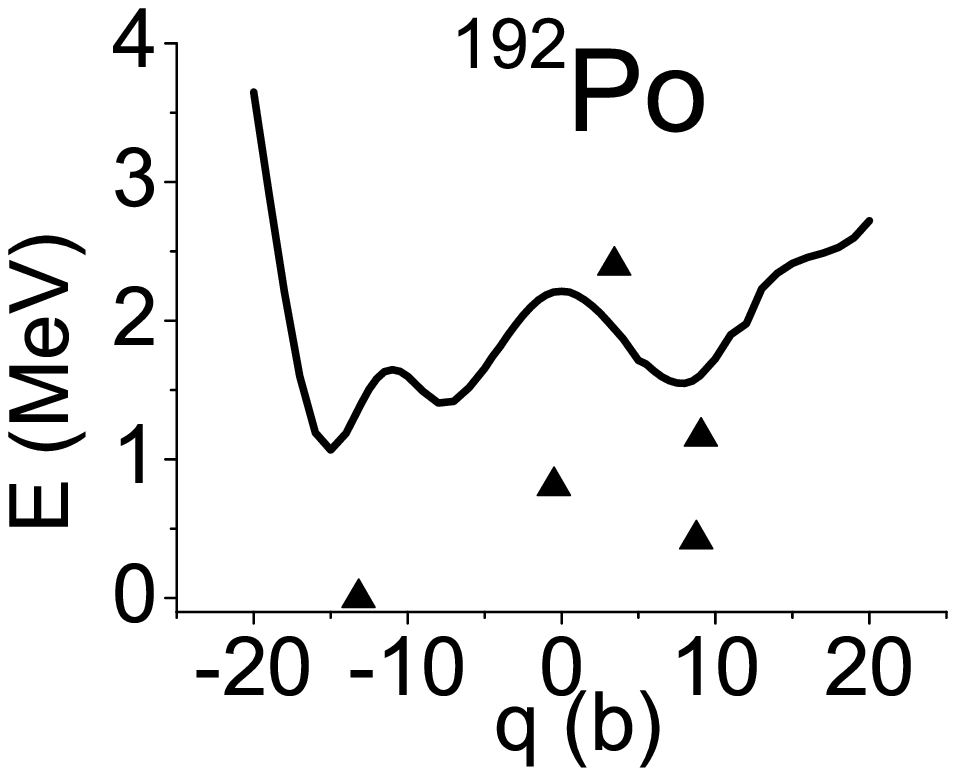} & 
 \includegraphics[height=.12\textheight]{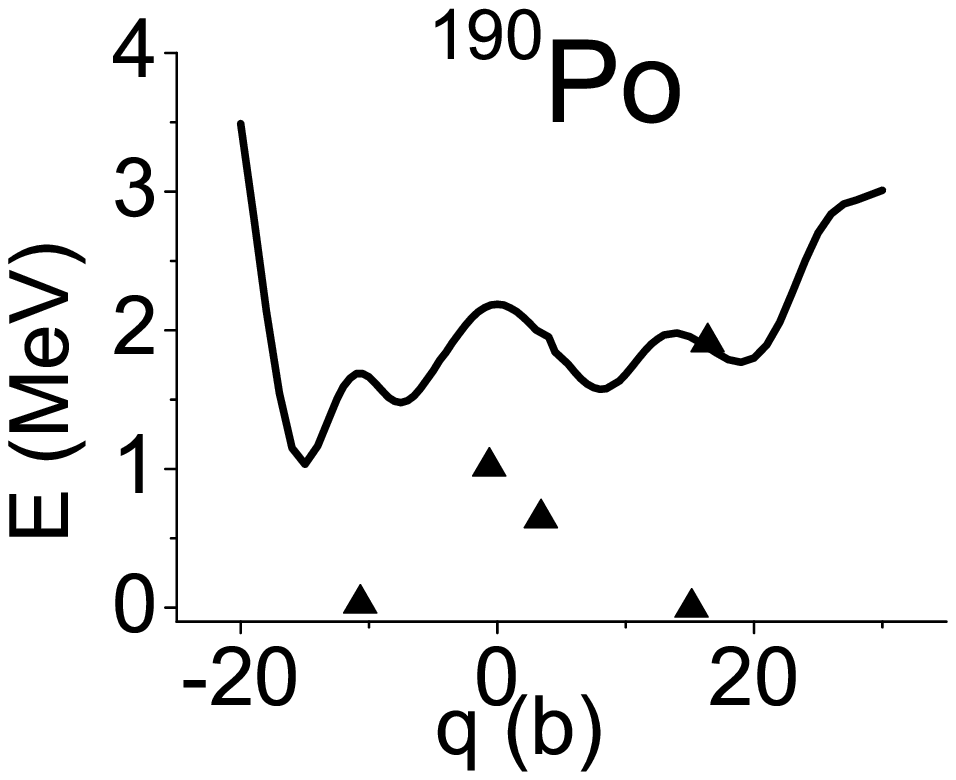} \\ 
 \includegraphics[height=.12\textheight]{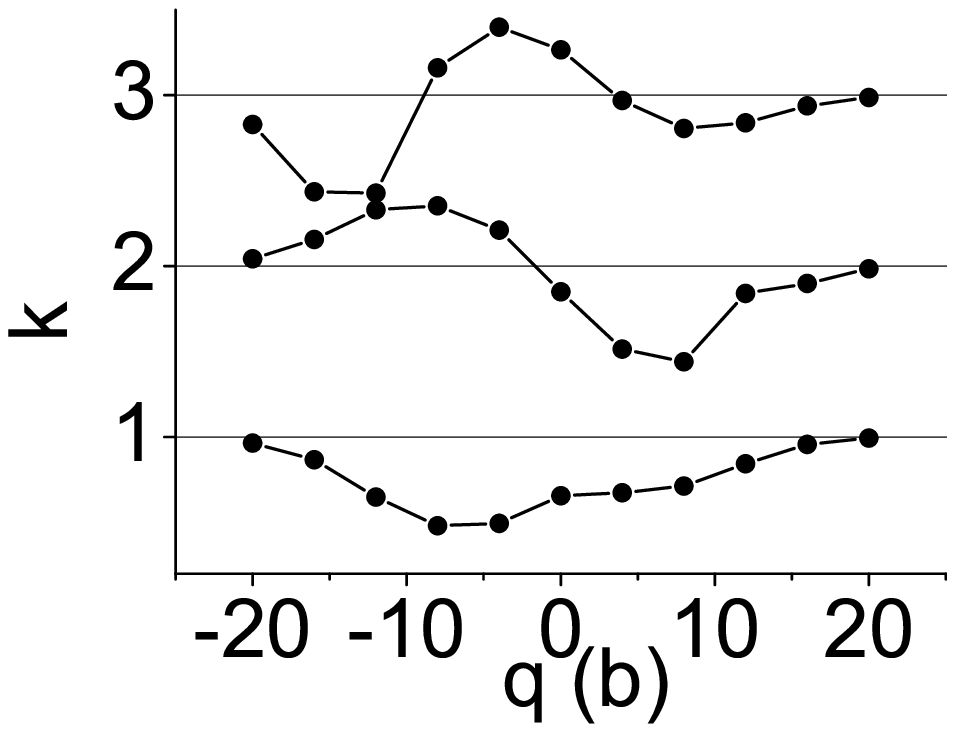}  & 
 \includegraphics[height=.12\textheight]{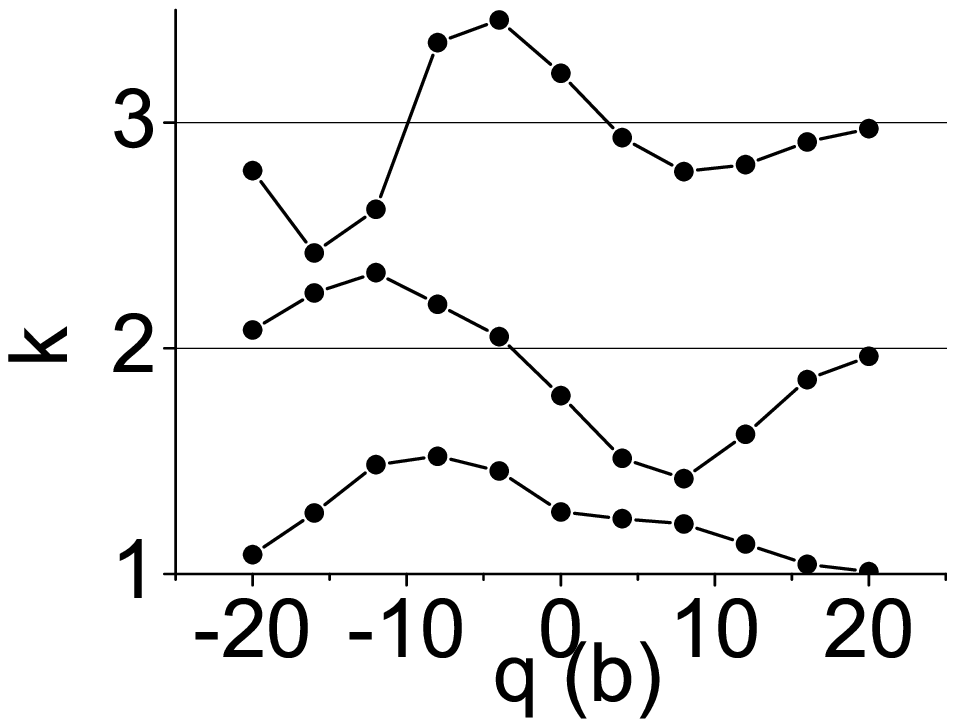}  & 
 \includegraphics[height=.12\textheight]{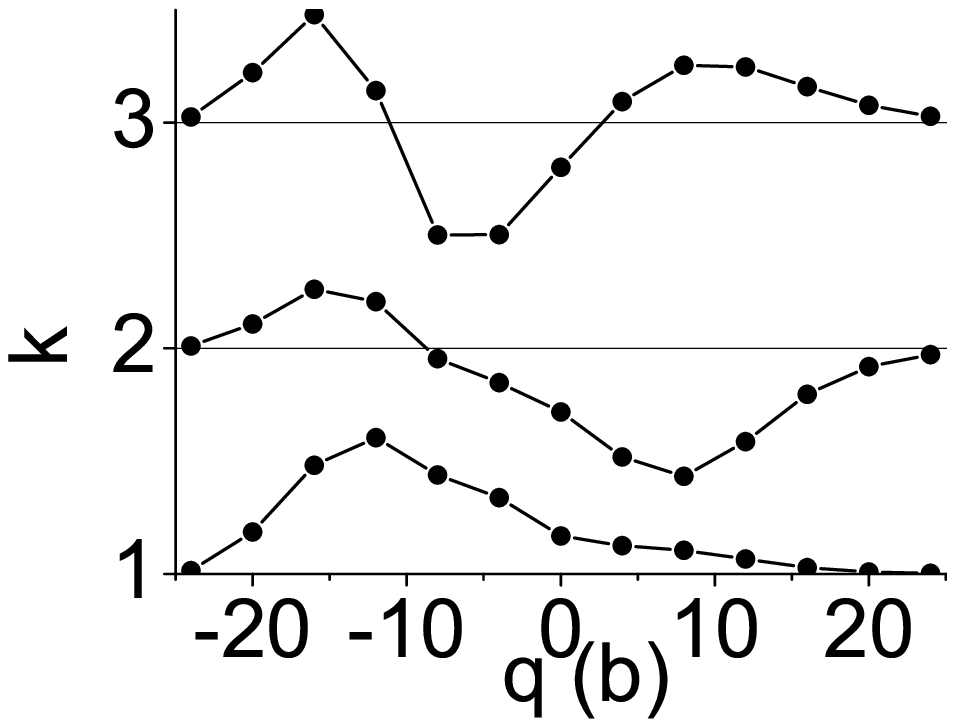}  & 
 \includegraphics[height=.12\textheight]{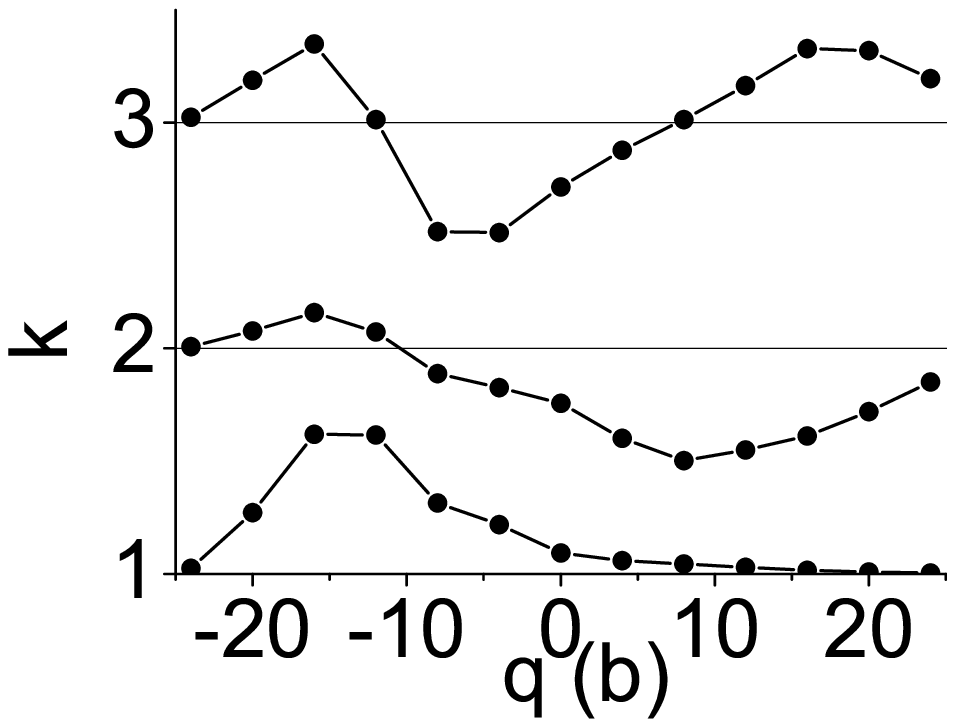}  & 
 \includegraphics[height=.12\textheight]{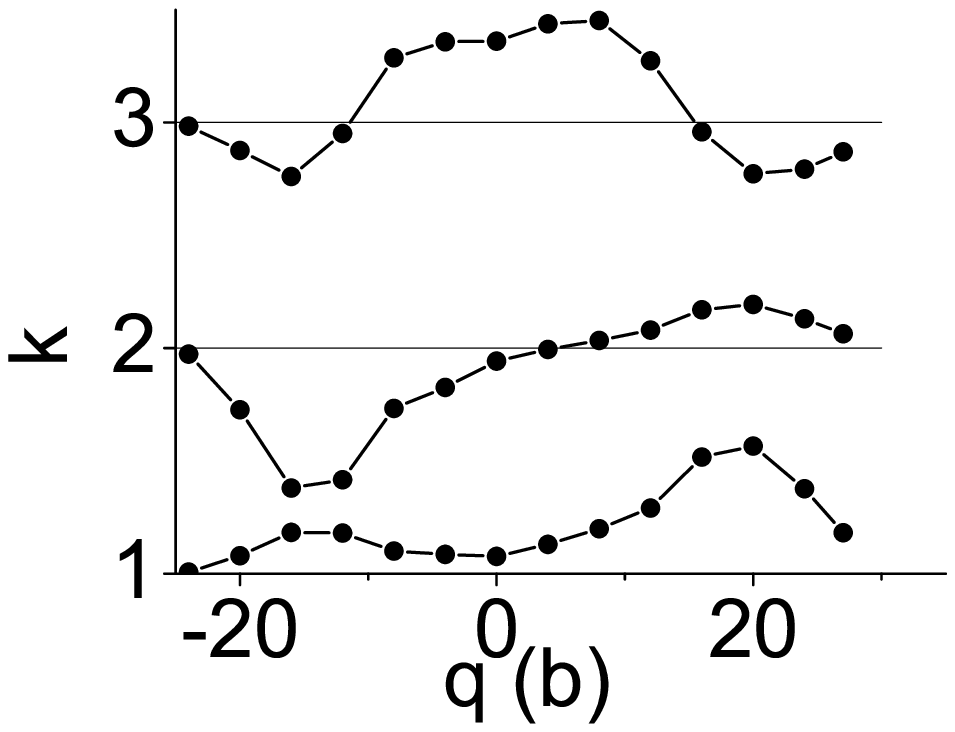} \\ 
\end{tabular} 
 \caption{Energies of the GCM eigenstates, together with 
the HF+BCS deformation energy curves for $^{198-190}$Po 
 (upper part). The amplitude of the corresponding  collective 
GCM eigenfunctions for three lowest states enumerated by $k$ 
are shown in the lower part of the figure.} 
\end{figure}

\begin{figure} 
 \includegraphics[height=.25\textheight]{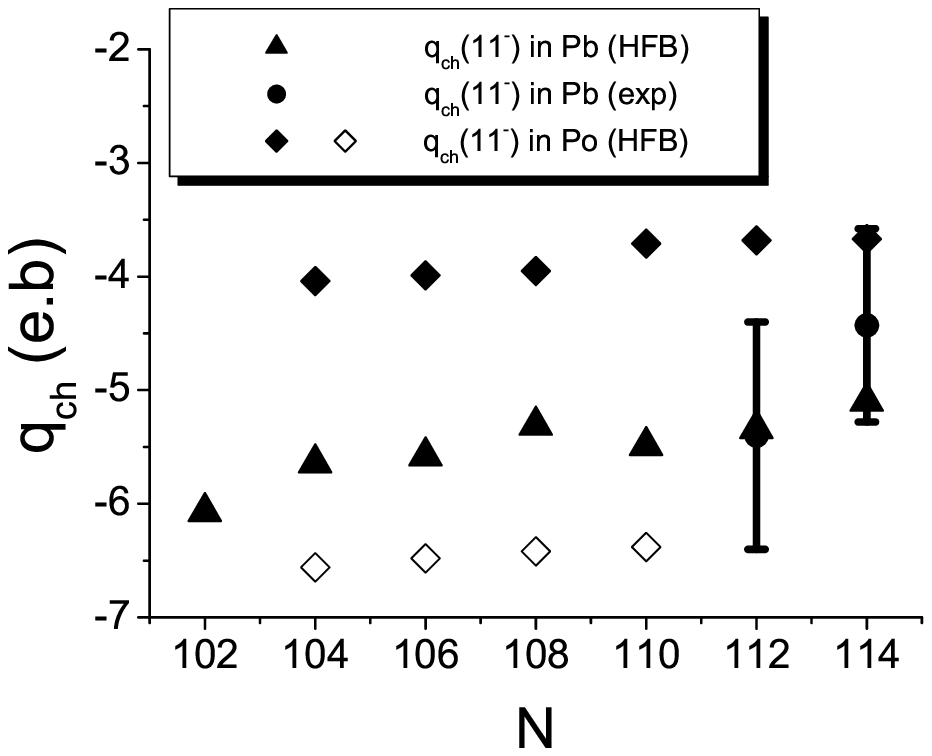} 
 \caption{Comparison between the charge quadrupole moment of the $11^-$ states in Pb 
   (triangles) and Po (diamonds). 
   The empty diamonds correspond to the $11^-$ states constructed on the 
   second oblate minima in light Po isotopes. 
   The intrinsic quadrupole moment as 
   extracted from the experiment~\protect\cite{VyOr02,VyCh02} for 
   $^{196,194}$Pb are shown by circles with error bars.} 
\end{figure}


\begin{thebibliography}{11} 
 
\bibitem{Heyde83} K.~Heyde et al, Phys. Rep. 102 (1983) 291. 
 
\bibitem{Wood92} For a review see J.L. Wood et al,  Phys. Rep. 215 (1992) 101. 
 
\bibitem{AFN90} S.~\AA berg, H.~Flocard, W.~Nazarewicz, 
Annu. Rev. Nucl. Part. Sci. 40 (1990) 439. 
 
\bibitem{JK91} R.V.F. Janssens and T.L. Khoo, 
Annu. rev. Nucl. Part. Sci. {\bf 41} (1991) 301 
 
 
\bibitem{VDCo} P. Van Duppen et al, Phys. Rev. Lett. 52 (1984) 1974; 
P. Van Duppen et al, Phys. Lett. B 154 (1985) 354. 
 
\bibitem{BijnAnd} N.~Bijnens et al, Z. Phys. A 356 (1996) 3; 
A.~N.~Andreyev et al, J. Phys. G 25 (1999) 835. 
 
 
\bibitem{Andrei00} A.~N.~Andreyev et al, Nature (London) 405 (2000) 430. 
 
\bibitem{AndreiEPJ99} A.~N.~Andreyev et al, Eur. Phys. J. A 6 (1999) 381. 
 
\bibitem{HeJo87} K.~Heyde et al, Nucl. Phys. A 466 (1987) 189. 
 
\bibitem{HeVI94}  K.~Heyde, P.~Van~Isacker, J.L.~Wood, Phys. Rev. C 49 (1994) 559. 
 
 
\bibitem{BeNa89} R.~Bengtsson, W.~Nazarewicz, Z. Phys. A 334 (1989) 269. 
 
\bibitem{Naz93} W.~Nazarewicz, Phys. Lett. B 305 (1993) 195. 
 
 
\bibitem{JuHe01} R. Julin, K. Helariutta, M. Muikku, J. Phys. G 27 (2001) R109. 
 
\bibitem{PeHe87} J.~Penninga et al, Nucl. Phys. A 471 (1987) 535. 
 
\bibitem{DrBy02} G.D.~Dracoulis et al, Proc. Int. Conf. ``Frontiers of Nuclear 
  Structure'', Berkeley, 2002, edt. P.Fallon. 
 
\bibitem{ClWa93} R.M.~Clark et al, Nucl. Phys. A 562 (1993) 121. 
 
\bibitem{DrBy98} G.D.~Dracoulis et al, Phys. Lett.  B 432 (1998) 37. 
 
\bibitem{VyCh02} K.~Vyvey et al., Phys. Rev. Lett. 88 (2002) 102502. 
 
\bibitem{VyOr02} K.~Vyvey et al., Phys. Lett. B 538 (2002) 33. 
 
 
\bibitem{BijnPRL} N.~Bijnens et al, Phys. Rev. Lett. 75 (1995) 4571. 
 
\bibitem{DCDe00} C.~De Coster, B.~Decroix, K.~Heyde, Phys. Rev. C 61 (2000) 067306. 
 
\bibitem{FoHe02} R.~Fossion, K.~Heyde, G. Thiamova, P. Van Isacker, 
                 Phys. Rev. C 67 (2003) 024306. 
 
\bibitem{MaPa77} F.~R.~May, V.~V.~Pashkevich, S.~Frauendorf, 
                 Phys. Lett. B 68  (1977) 113. 
 
\bibitem{TaFl93} N.~Tajima, H.~Flocard, P.~Bonche, J.~Dobaczewski, 
  P.-H.~Heenen, Nucl. Phys. A 551 (1993) 409. 
 
\bibitem{ChEg01} R.R. Chasman, J.L. Egido, L.M. Robledo, 
                 Phys. Lett. B 513 (2001) 325. 
 
\bibitem{HeVa01} P.-H. Heenen, A. Valor, M. Bender, P. Bonche, H. Flocard, 
                 Eur. Phys. J. A 11 (2001) 393. 
 
\bibitem{DuBe02} T. Duguet, M.~Bender, P.~Bonche, P.-H.~Heenen, 
                 Phys. Lett B 559 (2003) 201. 
 
 
\bibitem{OrHe99} A.M.~Oros et al., Nucl. Phys. A 645 (1999) 107. 
 
 
\bibitem{ClMa00} R.M.~Clark, A.O.~Macchiaveli, 
                 Annu. Rev. Nucl. Part. Sci. 50 (2000) 1. 
 
\bibitem{Sly4} E.~Chabanat, P.~Bonche, P.~Haensel, R.~Schaeffer, 
               Phys. Scripta T  56 (1995) 231. 
 
\bibitem{RiBo99} C.~Rigollet, P.~Bonche, H.~Flocard, P.-H.~Heenen, 
                 Phys. Rev. C 59 (1999) 3120. 
 
\bibitem{TeHe96} J.~Terasaki, P.-H.~Heenen, H.~Flocard, P.~Bonche, 
                 Nucl. Phys. A 600 (1996) 371. 
 
 
\bibitem{GaBo94} B.~Gall, P.~Bonche, J.~Dobaczewski, H.~Flocard, P.-H.~Heenen, 
                 Z. Phys. A 348 (1994) 183. 
 
\bibitem{LN} H.J.~Lipkin, Ann. Phys. (N.Y.) 9 (1960) 272; 
                 Y.~Nogami, Phys. Rev. B 134 (1962) 313; 
              H.C.~Pradhan, Y.~Nogami, J.~Law, 
             Nucl. Phys. A 201 (1972) 357. 
 
\bibitem{HeJa98} P.-H.~Heenen, R.V.F.~Janssens, 
                 Phys. Rev. C 57 (1999) 159. 
 
\bibitem{TeFl96} J.~Terasaki, H.~Flocard, P.-H.~Heenen, P.~Bonche, 
                 Nucl. Phys. A 621 (1996) 706. 
 
\bibitem{TeFl97} J.~Terasaki, H.~Flocard, P.-H.~Heenen, P.~Bonche, 
                 Phys. Rev. C 55 (1997) 1231. 
 
\bibitem{DuBo01} T.~Duguet, P.~Bonche, P.-H.~Heenen, 
              Nucl. Phys. A 679 (2001) 427. 
 
 
\bibitem{BoDo90} P.~Bonche et al, Nucl. Phys. A 510 (1985) 466. 
 
\bibitem{HeBo93} P.-H.~Heenen, P.~Bonche, J.~Dobaczewski, H.~Flocard, 
Nucl. Phys. A 561 (1993) 367 . 
 
\bibitem{HiWh53} D.~L.~Hill, J.~A.~Wheeler, 
                 Phys. Rev. 89 (1953) 1102 
\bibitem{BHR03} M. Bender, P.-H. Heenen, and P.-G. Reinhard, 
                Rev. Mod. Phys {\bf 75} (2003) 121 
 
 
\bibitem{RS80} P.~Ring, P.~Schuck, {\em The Nuclear Many-Body Problem} 
 (Springer-Verlag, New York, 1980). 
 
\bibitem{HeCo99} K.~Helariutta et al, Eur. Phys. J. A 6 (1999) 289. 
 
\bibitem{Po190} K.~Van de Vel et al, Eur. Phys. J. A 17 (2003) 167. 
 
\bibitem{CiDo96} S.~Cwiok, J.~Dobaczewski, P.-H.~Heenen, P.~Magierski, W.~Nazarewicz, 
Nucl. Phys. A 611 (1996) 211. 
 
 
 
\end{thebibliography}
\end{document}